\documentclass[aps,amsmath,amssymb]{revtex4}

\usepackage{graphicx}
\usepackage{dcolumn}
\usepackage{bm}

\begin{document}


\title{Pair dynamics in the formation of molecules in a Bose-Einstein condensate}

\author{Pascal Naidon}
\email{pascal.naidon@lac.u-psud.fr}
\author{Fran\c{c}oise Masnou-Seeuws}
\affiliation{Laboratoire Aim\'e Cotton, CNRS, B\^at. 505 Campus d'Orsay, 91405 Orsay Cedex, France.}

\date{\today}

\begin{abstract}
We revisit the mean-field treatment of photoassociation and Feshbach resonances in a Bose-Einstein condensate previously used by various authors. Generalizing the Cherny and Shanenko approach ( Phys. Rev. E {\bf 62}, 1646-59 (2000) ) where the finite size of the potentials is explicitly introduced, we develop a two-channel model for a mixed atomic-molecular condensate. Besides the individual dynamics of the condensed and non-condensed atoms, the model also takes into account their pair dynamics by means of pair wave functions. We show that the resulting set of coupled equations can be reduced to the usual coupled Gross-Pitaevski\u{\i} equations when the time scale of the pair dynamics is short compared to that of the individual dynamics. Such time scales are discussed in the case of typical photoassociation experiments with cw lasers. We show that the individual dynamics plays a minor role, demonstrating the validity of the rates predicted by the usual models describing photoassociation in a nondegenerate gas.
   
\end{abstract}

\maketitle

\section{INTRODUCTION}

The possibility of transforming an atomic condensate into a  molecular condensate is presently a major challenge \cite{julienne98,timmermans99b,javanainen99b,heinzen00,claussen02,claussen02b} . Several routes are considered to couple a condensate of free atoms with a condensate of molecules which are all in the same vibrational state of a molecular electronic potential. A Feshbach resonance in the electronic ground state can be swept by a time-dependent magnetic field and recent experiments \cite{claussen02,claussen02b} have observed oscillations in the number of atoms in the condensate. Alternatively, the photoassociation process, which can be considered as an optically induced Feshbach resonance, creates a molecular condensate in an excited electronic state. In the latter case, a stabilization process must be introduced to avoid destruction of this condensate by spontaneous emission as was observed recently \cite{mckenzie02}.  A two-pulse STIRAP (stimulated Raman adiabatic passage) scheme has been theoretically discussed \cite{mackie00,hope01,drummond02} in view of transferring the population to bound levels in the ground electronic state. \\

In a nondegenerate gas, ultracold molecules are formed through combination of a photoassociation step with a stabilization step using  spontaneous emission \cite{fioretti98,nikolov00,gabbanini00}.  In both cases the efficiency is controlled by the dipole transition moments, which depend markedly upon details of the electronic potential curves : the search for efficient mechanisms relies upon accurate spectroscopic data \cite{bahns00,masnou01} and it was shown that whereas photoassociation is efficient at large internuclear distances, the stabilization process is governed by the probability of finding the two atoms at intermediate distances\cite{masnou01,dulieu03}. Making ultracold molecules thus involves an interplay between long-range and short-range dynamics. Up to now, most experiments have used cw lasers. The use of chirped pulses, {\it ie} laser pulses with a time-dependent frequency, could increase the formation rates \cite{vala01}. From the theoretical point of view, sweeping a time-dependent magnetic field through a Feshbach resonance is equivalent to photoassociating with a chirped pulse. 

In a condensate, most theoretical treatments at present rely upon coupled Gross-Pitaevski\u{\i} equations, where the dynamics in the atomic and the molecular condensates as well as the coupling between them, are described by mean-field effective potentials \cite{timmermans99a,javanainen99b,drummond02}. Details of the potentials are omitted owing to a delta function approximation (contact potential). The knowledge of the molecular potential curves and dipole transition moments is necessary only to determine scattering lengths, binding energies and to accurately compute the coupling parameter between atomic and molecular phases. The success of such calculations relies on the very short time scale of the microscopic dynamics compared to the condensate dynamics.\\

The validity of the (one-body) mean-field approximation has been recently questioned by several authors, particularly in the case of a time-dependent coupling term, and models using Hartree-Fock-Bogoliubov equations of motion have been proposed  \cite{kokkelmans02}. However, such calculations rely on the delta function approximation for the potentials and coupling terms. The ultraviolet divergence caused by this approximation is solved by a renormalization procedure, as discussed in detail by Kokkelmans et al. \cite{kokkelmans02b}. Recently, Cherny and Shanenko \cite{cherny00,cherny00b} have shown that in the description of the dynamics of an atomic condensate, issues associated with the contact potentials can be avoided by considering the exact potential and pair wave functions having the correct nodal structure at short interatomic distance.

The aim of the present work is to revisit the problem of coupled atomic and molecular condensates in the framework of a Cherny treatment. For the sake of clarity, we shall consider only a two-channel model for Feshbach resonance or photoassociation.
The paper is organised as follows. We first present a three-field model describing a two-channel coupling in a Bose system. We then derive one-body and two-body mean-field equations, and show how the pair wave functions can be eliminated and lead to effective one-body mean-field equations, without using contact potentials. In the last section, we interpret the one-colour photoassociation of a BEC in terms of one-body and two-body modes. Definitions of the pair wave functions in an inhomogeneous system are given in the Appendix.

\section{THREE-FIELD MODEL}

\subsection{Many-body Hamiltonian}

\begin{figure}[t]
  \begin{center}
    $(a)$\includegraphics[width=6cm]{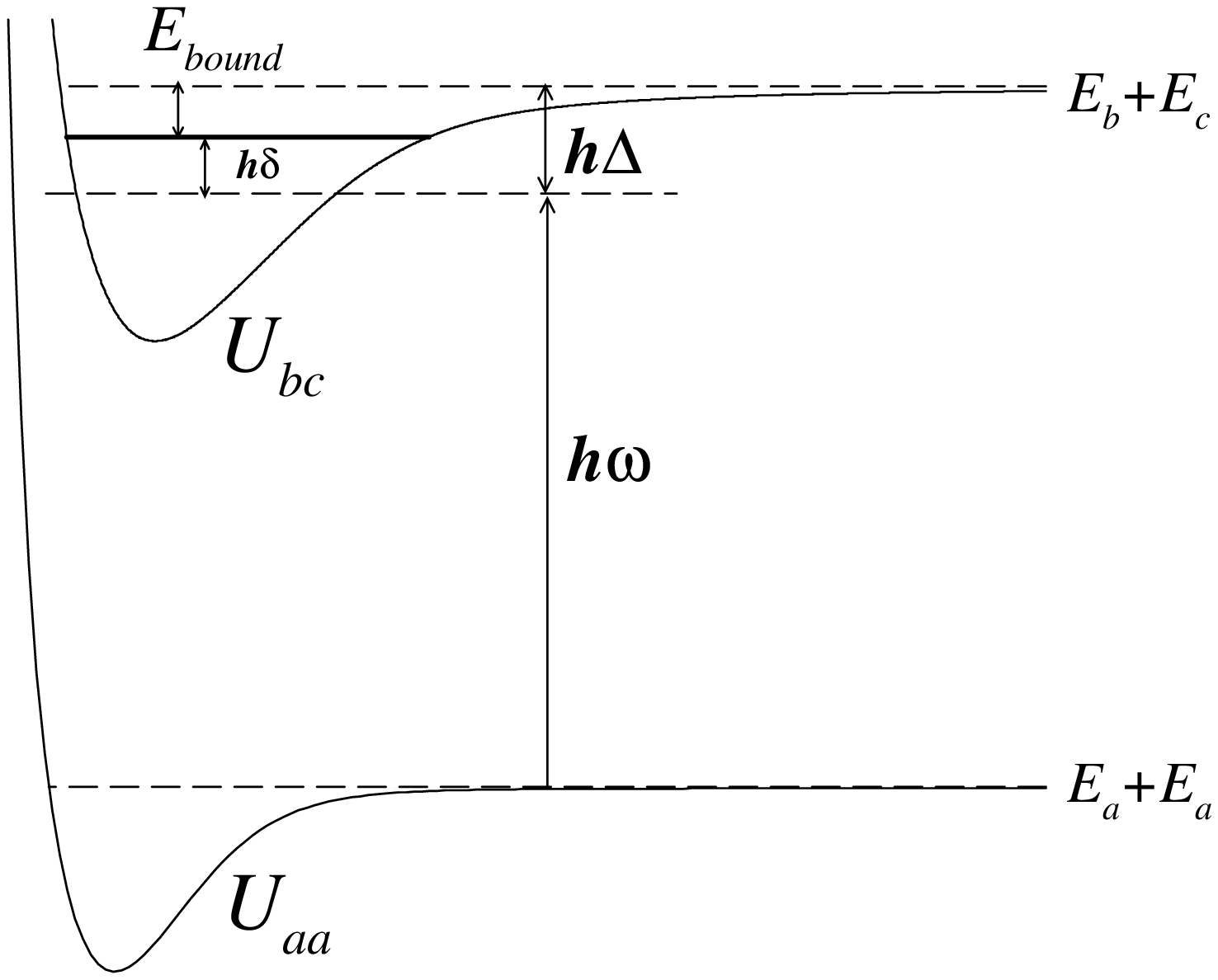}
    $(b)$\includegraphics[width=6cm]{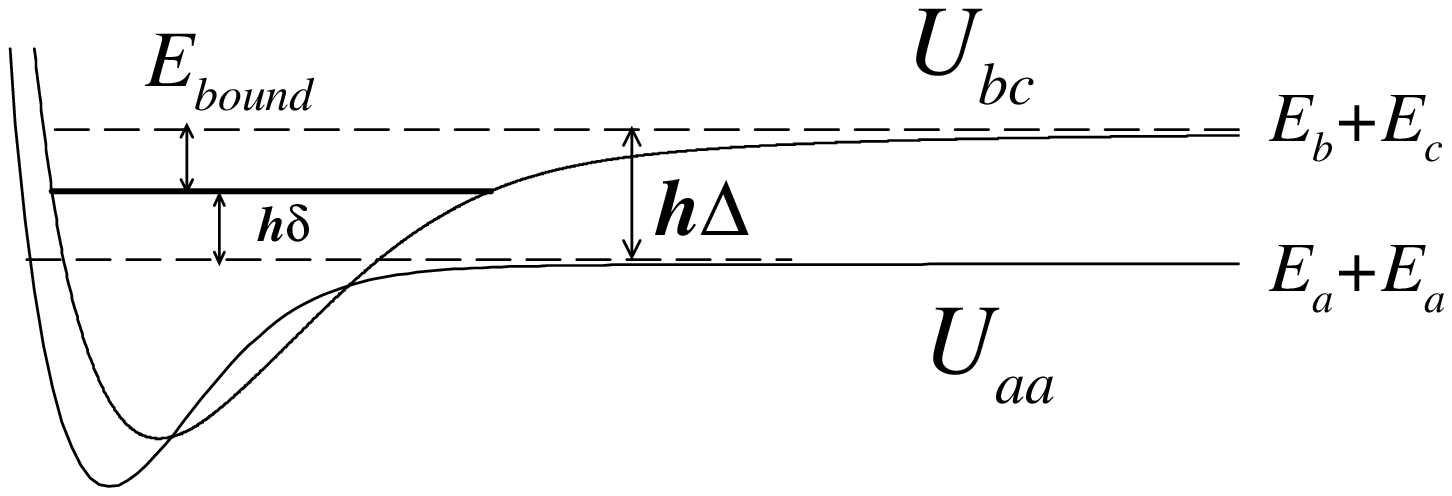}
  \end{center}
  \caption{
    Schematic representations of two coupled channels. $(a)$ coupling with a photoassociation laser of frequency $\omega/2\pi$: the asymptotic separation between the dressed potentials is given by a detuning $\Delta = E_b+E_c-2E_a-\hbar\omega$. $(b)$ Feshbach resonance: a magnetic field is adjusted to couple the two channels by hyperfine interaction. The potentials are separated asymptotically by a detuning $\Delta = E_b+E_c-2E_a$.
  }
  \label{fig:channels}
\end{figure}

 We consider three kinds of atoms in the atomic-molecular system: the ground-state atoms ($a$), colliding in the lower channel and the atoms ($b$ and $c$) of the bound pairs in the upper channel. As described in Figs.\ref{fig:channels}, the free atoms $a$, interact through the potential $U_{aa}(r)$, while the molecules vibrate in the potential $U_{bc}(r)$. In the case of photoassociation, the latter potential corresponds to an excited electronic state of the molecule. For each species, we define a quantum field: $\hat{\psi}_a$, $\hat{\psi}_b$, and $\hat{\psi}_c$.

This description of the system is a phenomenological starting point. It would be rigorous if the states $a$, $b$, and $c$ corresponded to well-defined atomic internal states, which is true only at large interatomic distances. Indeed, when two atoms come close to each other, their internal states change because their electronic clouds overlap. This means that, for instance, the ket $\frac{1}{\sqrt 2}\hat{\psi}^\dag_a({\bf x})\hat{\psi}^\dag_a({\bf y})|0\rangle$ is related to the ``molecular'' ket $\frac{1}{\sqrt 2} (|{\bf x},{\bf y} \rangle + |{\bf y},{\bf x} \rangle)\otimes|\phi^{el}_{aa}\rangle$, where $|\phi^{el}_{aa}\rangle$ is the molecular electronic ket depending on the distance $|{\bf x} - {\bf y}|$.

Bearing in mind this phenomenological aspect, we can write the many-body Hamiltonian of the system in terms of $\hat{\psi}_a$, $\hat{\psi}_b$, and $\hat{\psi}_c$:
\begin{eqnarray}
  \hat{H} &= &\sum_{i=a,b,c}
    \int\! {
      d^3\! {\bf x} ~ \hat{\psi}^\dag_i({\bf x})
      \left (
        -\frac{\hbar^2 \nabla^2}{2m}
        +V({\bf x}) + E_i
      \right )
      \hat{\psi}_i({\bf x})
    }
    \nonumber
  \\
  &+&
    \int\!\!\!\!\int\! {
      d^3\! {\bf x} d^3\! {\bf y} ~
      \frac{\hat{\psi}^\dag_a({\bf x}) \hat{\psi}^\dag_a({\bf y})}{\sqrt 2}
      U_{aa}({\bf x}\!-\!{\bf y})
      \frac{ \hat{\psi}_a ({\bf y}) \hat{\psi}_a({\bf x}) }{\sqrt 2}
    }
    \nonumber
  \\
   &+&
    \int\!\!\!\!\int\! {
      d^3\! {\bf x} d^3\! {\bf y} ~
      \hat{\psi}^\dag_b({\bf x}) \hat{\psi}^\dag_c({\bf y})
      U_{bc}({\bf x}\!-\!{\bf y})
      \hat{\psi}_c ({\bf y}) \hat{\psi}_b({\bf x})
    }
    \nonumber
    \\
       \label{Hamiltonian}
  &+&   \int\!\!\!\!\int\! {
         d^3\! {\bf x} d^3\! {\bf y} ~
          \hat{\psi}^\dag_b({\bf x}) \hat{\psi}^\dag_c({\bf y})
         H_{int}({\bf x},{\bf y}) 
         \frac{\hat{\psi}_a({\bf y}) \hat{\psi}_a({\bf x})}{\sqrt 2}                   
      }+h.c.
\end{eqnarray}

where $m$ is the mass of the atoms and $E_a$, $E_b$, $E_c$ are the internal energies of isolated atoms (see Fig.\ref{fig:channels}). $V$ is the potential trapping the atoms, $U_{aa}$ ($U_{bc}$) is the interaction potential between atoms of the lower (upper) channel, and $H_{int}$ is the matrix element coupling the two channels (which can be time-dependent). No specific approximation is made regarding these potentials, so that the Hamiltonian is built on microscopic grounds. In this respect, our approach is very close to that of \cite{kohler02}, the major difference being that we consider the two channels explicitly. Note that there are no terms involving the potentials $U_{ab}$, $U_{ac}$, $U_{bb}$, etc: since the atoms $b$ and $c$ are bound, we neglect their collisions with other atoms. Moreover, three-body potentials, as well as spontaneous emission (in the case of photoassociation), are not taken into account in this Hamiltonian.

\subsection{Dynamics of a two-atom system}
One can easily derive the usual dynamics of a two-atom system from the many-body Hamiltonian. For instance, consider the two-body state:
\begin{equation}
|\Omega\rangle = 
      \int\!\!\!\!\int
        d^3\! {\bf x} d^3\! {\bf y} ~ 
        {\Big (}
        \Phi_A ({\bf x},{\bf y}, t) \frac{ \psi^\dag_a({\bf x}) \psi^\dag_a({\bf y}) }{\sqrt 2}
        +~
        \Phi_M ({\bf x}, {\bf y}, t) \psi^\dag_b({\bf x}) \psi^\dag_c({\bf y}) 
        {\Big )}
  |0\rangle
  \nonumber
\end{equation}
where $\Phi_A$ and $\Phi_M$ are the two components of the two-body wave function for the lower and upper channels. As we will consider continuum states in the lower channel and bound states in the upper channel, $A$ stands for ``atomic'' and $M$ for ``molecular''. This state evolves according to the Schr\"odinger equation $ i\hbar \frac{d |\Omega\rangle }{dt} = \hat{H} | \Omega \rangle $. Using the canonical commutation relations $[\hat\psi_i({\bf x}),\hat\psi_j^\dag({\bf y})]=\delta_{ij}\delta^3({\bf x\!-\!y})$ and $[\hat\psi_i({\bf x}),\hat\psi_j({\bf y})]=0$, one finds a set of equations describing a general two-channel coupling:
\begin{widetext}
\begin{eqnarray}
\label{twoBodyDyn1}
  i\hbar \frac{d \Phi_A}{dt} &=&
  \left (
    -\frac{\hbar^2 (\nabla^2_x\!+\!\nabla^2_y)}{2m} + V({\bf x})+ V({\bf y})
    +U_{aa}({\bf x\!-\!y}) + 2E_a
  \right )
  \Phi_A
  ~+~
  H^*_{int}({\bf x},{\bf y})\Phi_M
\\
\label{twoBodyDyn2}
  i\hbar \frac{d \Phi_M}{dt} &=&
  \left (
    -\frac{\hbar^2 (\nabla^2_x\!+\!\nabla^2_y)}{2m} + V({\bf x})+ V({\bf y})
    +U_{bc}({\bf x\!-\!y}) + E_b + E_c
  \right )
  \Phi_M
  \nonumber
\\
&&
  ~+~
  H_{int}({\bf x},{\bf y})\Phi_A
\end{eqnarray}
\end{widetext}

Usually, one can separate the motion of the centre of mass ${\bf R}=\frac{\bf x+y}{2}$ and only the relative coordinate ${\bf r}={\bf x}-{\bf y}$ is considered:
\begin{eqnarray}
\label{RelativeDyn1}
  i\hbar \frac{d \Phi_A ({\bf r}, t)}{dt} &=&
  \left (
    -\frac{\hbar^2 \nabla^2_r}{m}
    +U_{aa}({\bf r}) + 2E_a
  \right )
  \Phi_A({\bf r}, t)
  ~+~
  H^*_{int}({\bf r})\Phi_M({\bf r},t)
\\
\label{RelativeDyn2}
  i\hbar \frac{d \Phi_M ({\bf r}, t)}{dt} &=&
  \left (
    -\frac{\hbar^2 \nabla^2_r}{m}
    +U_{bc}({\bf r}) + E_b + E_c
  \right )
  \Phi_M({\bf r}, t)
  ~+~
  H_{int}({\bf r})\Phi_A({\bf r},t)
\end{eqnarray}

\subsection{Dynamics of a many-atom system}
In the more general case of a many-body state, the full dynamics is given by the equations of motion for the field operators in the Heisenberg picture. These equations are obtained from the Heisenberg equations $i\hbar \frac{d \hat{O}}{dt} = [\hat{O}, \hat{H}]$, using the canonical commutation relations. 
In the case of photoassociation with a continuous laser, we may actually rotate the field operators and use the rotating field approximation in order to remove the oscillatory time dependence of $H_{int}$ \cite{mihaela01}. In either case, we find:
\begin{eqnarray}
i\hbar \frac{d \hat{\psi}_a ({\bf x})}{dt} &=&
\left (
  -\frac{\hbar^2 \nabla^2_x}{2m} + V({\bf x}) + \int\!\! d^3\! {\bf y}~ \underline{U_{aa}({\bf x}\!\!-\!\!{\bf y})} \hat{\psi}^\dag_a({\bf y}) \hat{\psi}_a({\bf y})
\right ) \hat{\psi}_a ({\bf x})
\nonumber
\\
&&+ \sqrt 2 \int\!\! d^3\! {\bf y}~ \hat{\psi}^\dag_a({\bf y}) \underline{H_{int}^*({\bf x},{\bf y}) \hat{\psi}_b({\bf x}) \hat{\psi}_c({\bf y})}
\label{operatorDyn_a}
\\
  i\hbar \frac{d \hat\psi_b({\bf x})\hat\psi_c({\bf y})}{dt} &=&
  \left (
    -\frac{\hbar^2 (\nabla^2_x\!+\!\nabla^2_y)}{2m} + V({\bf x})+ V({\bf y})
    +U_{bc}({\bf x\!-\!y}) + \hbar\Delta
  \right )
  \hat\psi_b({\bf x})\hat\psi_c({\bf y})
\nonumber
\\
  && ~+~ H_{int}({\bf x},{\bf y})\frac{\hat\psi_a({\bf x})\hat\psi_a({\bf y})}{\sqrt2}
\label{operatorDyn_bc}
\end{eqnarray}

where underlining has been used as a notational convenience for symmetrizing certain quantities; for instance, $\underline{A({\bf x},{\bf y})}$ actually means $\frac{1}{2}(A({\bf x},{\bf y})+A({\bf y},{\bf x}))$. In Eq. (\ref{operatorDyn_bc}), we introduced the ``detuning'' $\Delta$ between the two asymptotic curves (see Fig.\ref{fig:channels}) and we neglected many-body terms corresponding to collisions or coupling with atoms external to the pair considered. Keeping these terms would be inconsistent with the fact that we neglected other potentials such as $U_{ab}$, $U_{ac}$.

\subsection{Coupling with a single bound-state}
Usually, the interaction is tuned to couple the atoms in the ground-state channel to a single stationary bound state in the upper channel: a rovibrational level of the potential $U_{bc}$ with binding energy $E_{bound}$.  The relative motion of the bound atoms is then described by the stationary wave function $\varphi_M$ satisfying:
\begin{equation}
\label{varphi_M}
\Big( -\frac{\hbar^2\nabla_r^2}{m} + U_{bc}({\bf r}) \Big) \varphi_M({\bf r}) = -E_{bound}\varphi_M({\bf r})
\end{equation}

Here the zero of energy is set to $E_b + E_c$. To achieve population of this single level, one must remain in the perturbative regime, where the typical intensity of the coupling $\bar H_{int}$ remains smaller than the energy splittings between the molecular levels. This condition has been discussed in detail by Vatasescu {\it et al} \cite{mihaela01}, in the case of photoassociation in a trap of cold alkali atoms, considering various c.w laser intensities and detunings $\Delta$ and comparing the Rabi period to the classical vibrational period of the molecular level.\\

When $\varphi_M$ is indeed the only bound level affected by the coupling, we can approximate the operator ${\hat\psi}_b({\bf x}){\hat\psi}_c({\bf y})$ by its projection ${\hat\psi}_M(\frac{\bf x\!+\!y}{2}) \varphi_M({\bf x\!-\!y})$ onto this bound level, where the ``time-dependent coefficient'' ${\hat\psi}_M({\bf R}) = \int\! d^3{\bf r}~\varphi^*_M({\bf r}){\hat\psi}_b({\bf R}+\frac{\bf r}{2}){\hat\psi}_c({\bf R}-\frac{\bf r}{2})$ defines a molecular field. The description of the system can then be reduced to two fields $\hat{\psi}_a$ and $\hat{\psi}_M$, satisfying:
\begin{eqnarray}
i\hbar \frac{d \hat{\psi}_a ({\bf x})}{dt} &=&
\left (
  -\frac{\hbar^2 \nabla^2_x}{2m} + V({\bf x}) + \int\!\! d^3\! {\bf y}~ \underline{U_{aa}({\bf x}\!\!-\!\!{\bf y})} \hat{\psi}^\dag_a({\bf y}) \hat{\psi}_a({\bf y})
\right ) \hat{\psi}_a ({\bf x})
\nonumber
\\
&&~+~\sqrt 2 \int\!\! d^3\! {\bf y}~ \underline{W^*({\bf x, y})} \hat{\psi}^\dag_a({\bf y}) \hat{\psi}_M(\frac{\bf x\!+\!y}{2})
\label{TwoFieldDyn_a}
\\
  i\hbar \frac{d \hat\psi_M({\bf x})}{dt} &=&
  \left (
    -\frac{\hbar^2 \nabla^2_x}{4m} + 2V({\bf x})
    +\hbar\delta
  \right )
  \hat\psi_M({\bf x})
\nonumber
\\
  &&~+~
  \int\!\! d^3\! {\bf r}~W({\bf x}\!+\!\frac{\bf r}{2},{\bf x}\!-\!\frac{\bf r}{2})\frac{\hat\psi_a({\bf x}\!+\!\frac{\bf r}{2})\hat\psi_a({\bf x}\!-\!\frac{\bf r}{2})}{\sqrt2}
\label{TwoFieldDyn_M}
\end{eqnarray}

where $W({\bf x, y}) = H_{int}({\bf x},{\bf y}) \varphi_M^*({\bf x\!-\!y})$ is the interaction kernel and $\hbar\delta = \hbar\Delta - E_{bound}$ is the energy detuning between the two levels (see Figs. \ref{fig:channels}), considering the dressed picture for photoassociation. 

\section{EFFECTIVE MEAN-FIELD THEORY}
\subsection{Purely atomic system}

Let us first consider a purely atomic system (thus ignoring the terms involving any field $\hat{\psi}_b$ or $\hat{\psi}_c$). The equation of motion (\ref{operatorDyn_a}) now simply reads:
\begin{eqnarray}
  \label{PureAtomicMotion_a}
  i\hbar \frac{d \hat{\psi}_a ({\bf x})}{dt} &=&
  \left (
    -\frac{\hbar^2 \nabla^2_x}{2m} + V({\bf x}) + \int\! d^3\! {\bf y}~ \underline{U_{aa}({\bf x}\!-\!{\bf y})}
     \hat{\psi}^\dag_a({\bf y})\hat{\psi}_a({\bf y})
  \right ) \hat{\psi}_a ({\bf x})
\end{eqnarray}

In the condensed phase, one usually assumes a non-zero expectation value $\Psi_0({\bf x}) \equiv \langle \hat{\psi}_a({\bf x}) \rangle$ for the field operator $\hat{\psi}_a({\bf x})$, corresponding to a macroscopically occupied state (see the Appendix). What is sometimes referred to as ``naive mean-field'' consists in replacing the field operators by their averages directly in the equation of motion (\ref{PureAtomicMotion_a}), thus neglecting the quantum fluctuations $\hat\theta_a \equiv \hat\psi_a - \Psi_0$. This leads to a non-linear Schr\"{o}dinger equation with a coupling constant of the form $\int\!\! d^3\!{\bf x}~ U_{aa}({\bf x})$. This coupling constant is well defined for weak-coupling interactions (for instance, interatomic potentials satisfying the Born approximation $\int\!\! d^3\!{\bf x}~ U_{aa}({\bf x}) \approx \frac{4 \pi a \hbar^2}{m}$ where $a$ is the s-wave scattering length). It is not the case however for the interactions considered here. The interatomic potential exhibits a strong repulsive hard-core which leads to a divergent coupling constant.

 The usual remedy is to replace the real interaction by an effective one \cite{huang1987}, generally a contact potential $U_{aa}({\bf r}) = \frac{4 \pi a \hbar^2}{m}\delta^3({\bf r})$ (which does satisfy the Born approximation) having the same scattering length $a$. The resulting equation, known as the Gross-Pitaevski\u{\i} equation \cite{gross61,pitaevskii61}, forms an ``effective mean-field'' theory, in which only the large-scale effects of the interaction are retained:
\begin{equation}
  \label{GrossPitaevskii}
  i\hbar \frac{d \Psi_0 ({\bf x})}{dt} =
  \left (
    -\frac{\hbar^2 \nabla^2_x}{2m} + V({\bf x}) + \frac{4 \pi a \hbar^2}{m} | \Psi_0({\bf x}) |^2
  \right ) \Psi_0 ({\bf x})
\end{equation}

This method can be rigorously justified by considering the effect of two-body correlations \cite{proukakis97,cherny00}. An intuitive way of taking these correlations into account is to use pair wave functions. Just as most atoms are individually described by the same macroscopic wave function $\Psi_0({\bf x})$, most pairs of atoms are described by the same macroscopic pair wave function $\Phi_{00}({\bf x},{\bf y})$, which turns out to be simply the anomalous average $\frac{1}{\sqrt2} \langle \hat{\psi}_a({\bf x}) \hat{\psi}_a({\bf y})\rangle$ (See Eq. (\ref{Psi00}) in the Appendix). At large distances, the atoms are decorrelated, and the pair wave function is just a product of two macroscopic one-body wave functions $ \langle \hat{\psi}_a({\bf x}) \hat{\psi}_a({\bf y})\rangle \approx  \langle \hat{\psi}_a({\bf x}) \rangle \langle \hat{\psi}_a({\bf y})\rangle$. So we may write $\Phi_{00}({\bf x},{\bf y}) = \frac{1}{\sqrt2} \Psi_0({\bf y}) \Psi_0({\bf y})\varphi_{00}({\bf x},{\bf y})$ with $\lim_{|{\bf x}-{\bf y}| \rightarrow \infty} \varphi_{00}({\bf x},{\bf y}) = 1$. The function $\varphi_{00}$ may be seen as a reduced pair wave function describing the correlations at short distances (see Fig.\ref{fig:pair}).
To some approximations (see assumptions {\bf H1} and {\bf H2} in the Appendix) justified by the low density and large extent of the condensate, $\varphi_{00}$ simply satisfies the scattering equation for two atoms in free space. Written with the centre-of-mass and relative coordinates ${\bf R}$ and ${\bf r}$, this equation reads:
\begin{eqnarray}
  \label{Motion_varphi2}
  i\hbar \frac{d \varphi_{00} ({\bf R}, {\bf r},t)}{dt} &=~
  \Big (
    -\frac{\hbar^2 }{4m}\nabla^2_R
    -\frac{\hbar^2 }{m}\nabla^2_r
    +\underline{U_{aa}({\bf r})} 
  \Big ) ~
  \varphi_{00} ({\bf R}, {\bf r},t)
\end{eqnarray}
At equilibrium, it becomes a stationary state $\varphi^{(E)}$ satisfying :
\begin{equation}
\label{varphi^{(E)}}
  E \varphi^{(E)} ({\bf R}, {\bf r}) =
  \Big (
    -\frac{\hbar^2 }{4m}\nabla^2_R
    -\frac{\hbar^2 }{m}\nabla^2_r
    +\underline{U_{aa}({\bf r})} 
  \Big ) ~
  \varphi^{(E)} ({\bf R}, {\bf r})
\end{equation}
 
When the condensate pairs are in their ground state (which is presumably the case when the condensate is not excited), $\varphi_{00}$ is therefore the lowest energy state satisfying $\lim_{|{\bf r}| \rightarrow \infty} \varphi^{(E)}({\bf R},{\bf r}) = 1$, ie the stationary scattering state at zero energy $\varphi^{(0)}$ \footnote{The zero energy is a consequence of the assumption ${\bf H_1}$ that the condensate wave function is uniform over the scale of $\varphi$, and has therefore no momentum. In reality, the typical energies involved in current traps ($\sim nK$) lie in the threshold law regime}. This state can be expressed formally by:
\begin{equation}
\label{varphi_0}
  \varphi^{(0)}({\bf r}) = 1 - \int\!\! d^3\!{\bf R'} d^3\!{\bf r'}~\underline{U_{aa}({\bf r'})} G({\bf R, r, R', r'})
   = 1 - \int\!\! d^3\!{\bf r'}~\underline{U_{aa}({\bf r'})} g({\bf r, r'})
\end{equation}

where $G$ is the Green's function of the operator $-\frac{\hbar^2 }{4m}\nabla^2_R -\frac{\hbar^2 }{m}\nabla^2_r+\underline{U_{aa}({\bf r})}$ and $g$ the Green's function of the operator $-\frac{\hbar^2 }{m}\nabla^2_r+\underline{U_{aa}({\bf r})}$. It is known from collision theory \cite{joachain1983} that this state is related to the s-wave scattering length $a$:
\begin{equation}
  \label{scatteringLength}
  \int\!\! d^3\!{\bf r}~\underline{U_{aa}({\bf r})}~\varphi^{(0)}({\bf r}) = \frac{4\pi\hbar^2a}{m} \equiv g 
\end{equation} 

When taking the expectation value of the equation of motion (\ref{PureAtomicMotion_a}), one finds the term $\langle \hat{\psi}^{\dag}_a({\bf y}) \hat{\psi}_a({\bf y}) \hat{\psi}_a({\bf x}) \rangle$, which can be approximated by $\langle \hat{\psi}^{\dag}_a({\bf y}) \rangle \langle \hat{\psi}_a({\bf y}) \hat{\psi}_a({\bf x}) \rangle = |\Psi_0({\bf y})|^2\Psi_0({\bf x})\varphi_{00}({\bf x, y})$ when neglecting the collisions of non-condensed atoms with condensed atoms $(\rho a^3\ll1)$ \cite{cherny00}. If we assume again that the averaged field $\Psi_0$ is nearly uniform on the scale of $\varphi_{00}$, the interaction term becomes at equilibrium:
\begin{equation}
  \label{MeanFieldTerm}
  \left ( \int\! d^3{\bf y}~ \underline{U_{aa}({\bf x\!-\!y})}\varphi^{(0)}({\bf x\!-\!y}) \right ) |\Psi_0({\bf x})|^2 \Psi_0({\bf x})
  =
  \frac{4\pi\hbar^2 a}{m} |\Psi_0({\bf x})|^2 \Psi_0({\bf x})
\end{equation}

where we used (\ref{scatteringLength}). Thus we retrieve the mean-field term in the Gross-Pitavski\u{\i} equation (\ref{GrossPitaevskii}): it is in fact regularised by the stationary two-body correlations described by $\varphi^{(0)}$.

\subsection{Atomic and molecular system}

We now apply these ideas to the atomic-molecular system, with the aim of deriving an effective mean-field theory. A first approach \cite{timmermans99a,kostrun00,heinzen00} would be to start from the two-field description (\ref{TwoFieldDyn_a}),(\ref{TwoFieldDyn_M}) and replace the potentials by effective (or renormalised) interactions: $U_{aa}({\bf r}) = g \delta^3({\bf r})$ with $g = \frac{4 \pi  \hbar^2 a}{m}$ and $W({\bf r}) = w \delta^3({\bf r})$, with
\begin{equation}
 w = \sqrt2 \langle \varphi_M | H_{int} | \varphi^{(0)} \rangle \equiv \sqrt2 \int\!\!d^3\!{\bf r}~ \varphi_M^*({\bf r}) H_{int}({\bf r}) \varphi^{(0)}({\bf r})
\label{eq:w}
\end{equation}

Replacing the quantum fields by their averages $\langle \psi_a \rangle = \Psi_0$ and $\langle \psi_M \rangle = \Psi_M$, we obtain a set of coupled Gross-Pitaevski\u{\i} equations which has been extensively studied \cite{timmermans99a,kostrun00,heinzen00,drummond02,kostrun99,cusack01}:
\begin{eqnarray}
\label{MeanField10}
i\hbar \frac{d \Psi_0 ({\bf x})}{dt} &~=~&
\left (
  -\frac{\hbar^2 \nabla^2_x}{2m} + V({\bf x}) + g|\Psi_0({\bf x})|^2
\right ) \Psi_0 ({\bf x})
~~+~~ w^* \Psi^*_0({\bf x}) \Psi_M({\bf x})
\\
\label{MeanField1m}
i\hbar \frac{d \Psi_M ({\bf x})}{dt} &~=~&
\left (
  -\frac{\hbar^2 \nabla^2_x}{4m} + 2 V({\bf x}) + \hbar \delta
\right ) \Psi_M ({\bf x})
~~+~~ \frac{1}{ 2} w \Psi_0^2({\bf x})
\end{eqnarray}

Let us now follow the approach of \cite{cherny00}. We now consider the following one-body and two-body fields:
\begin{itemize}
\item the atomic condensate mode $\Psi_0$
\item the atomic non-condensed modes $\Psi_i$
\item the molecular condensed mode $\Phi_M = \langle\hat{\psi}_b\hat{\psi}_c\rangle$ (we will neglect non-condensed molecular modes)
\item the pair wave function  $\Phi_{00}$ for two condensed atoms
\item the pair wave function $\Phi_{0i}$ for a condensed atom and a non-condensed atom in the mode $\Psi_i$ (we will neglect the pairs of non-condensed atoms)
\end{itemize}

We first take the expectation value of Eqs. (\ref{operatorDyn_a}),(\ref{operatorDyn_bc}), expressing the atomic correlations by means of pair wave functions:
\begin{eqnarray}
i\hbar \frac{d \Psi_0 ({\bf x})}{dt} &=&
\Big (
  -\!\frac{\hbar^2 \nabla^2_x}{2m} + V({\bf x})
\Big ) {\Psi}_0 ({\bf x})
   +  \sqrt2\!
       \int\!\! d^3\! {\bf y}~
       \Big (
       \underline{U_{aa}({\bf x}\!\!-\!\!{\bf y})} \sum_{i}\Psi^{*}_i({\bf y}) \Phi_{0i}({\bf x},\!{\bf y}) 
\nonumber
\\
&& +~  \underline{H_{int}^*({\bf x},{\bf y})} \Psi^{*}_0({\bf y}) \Phi_M({\bf x},\!{\bf y})
       \Big )
\label{fieldDyn_0}
\\
  i\hbar \frac{d \Phi_M({\bf x},{\bf y})}{dt} &=&
  \Big (
    -\!\frac{\hbar^2 (\nabla^2_x\!+\!\nabla^2_y)}{2m} + V({\bf x})+ V({\bf y})
    +U_{bc}({\bf x\!-\!y}) + \hbar\Delta
  \Big )
  \Phi_M({\bf x},{\bf y})
\nonumber
\\
  &&+~
  H_{int}({\bf x},{\bf y})\Phi_{00}({\bf x},{\bf y})
\label{fieldDyn_M}
\end{eqnarray}
The dynamics of $\Psi_i$, $\Phi_{00}$ and $\Phi_{0i}$ is given in the Appendix. Note that Eqs. (\ref{Phi_00_evolution}) and (\ref{fieldDyn_M}) giving the evolution of $\Phi_{00}$ and $\Phi_M$ are analogous to Eqs. (\ref{twoBodyDyn1}),(\ref{twoBodyDyn2}). They describe indeed the coupled dynamics for the pairs of condensed atoms, giving rise to Rabi oscillations on a time scale
\begin{equation}
\label{T_pairs}
 T_{pairs}=\frac{\hbar}{\bar H_{int}}.
\end{equation}

To simplify these general equations, we make use of the reduced pair wave functions $\varphi_{00}$ and $\varphi_{0i}$ defined by Eqs.  (\ref{reduced_ii}),(\ref{reduced_ij}), with the assumptions ${\bf H_1}$ and ${\bf H_2}$.
Furthermore, we resort to a perturbative approach and assume that the atoms are coupled to a single stationary bound state $\varphi_M$, satisfying Eq. (\ref{varphi_M}). In this case, $\Phi_M({\bf x},{\bf y})$ reduces to its projection $\Psi_M(\frac{\bf x+y}{2})\varphi_M({\bf x\!-\!y})$ onto $\varphi_M$. The ``time-dependent coefficient'' $\Psi_M$ is the centre-of-mass wave function for the molecules in this bound state.

These assumptions lead to a closed set of equations:
\begin{eqnarray}
  \label{fieldDyn2_0}
  i\hbar \frac{d \Psi_0}{dt} &=&
  \big (
    -\frac{\hbar^2 \nabla^2}{2m} + V  +  g_{00} |\Psi_0|^2  +   \sum_{i\ne0} 2 g_{0i} |\Psi_i|^2
  \big ) \Psi_0
  ~+~   {g'}^*_M \Psi_M\Psi^*_0
\\
  \label{fieldDyn2_i}
  i\hbar \frac{d \Psi_i}{dt} &=&
  \big (
    -\frac{\hbar^2 \nabla^2}{2m} + V +   2 g_{0i} |\Psi_0|^2
  \big ) \Psi_i
\\
  \label{fieldDyn2_M}
  i\hbar \frac{d \Psi_M}{dt} &=&
  \Big (
    -\frac{\hbar^2 \nabla^2}{4m} + 2V
    + \hbar\delta
  \Big )
  \Psi_M
  ~+~
  \frac{1}{2}g_M
  \Psi_0^2
\\
  \label{fieldDyn2_00}
  i\hbar \frac{d\varphi_{00}({\bf R, r})}{dt}
  &=&
  \Big (
    -\frac{\hbar^2 \nabla^2_R}{4m}
    -\frac{\hbar^2 \nabla^2_r}{m}
    +\underline{U_{aa}({\bf r})}
  \Big )
   \varphi_{00}({\bf R, r}) \nonumber
\\~&~&   
  ~+~
  \sqrt2 H^*_{int}({\bf R,r}) \varphi_M({\bf r})
  \Big (
    \frac{\Psi_M({\bf R})}{\Psi_0({\bf R}\!+\!\frac{\bf r}{2})\Psi_0({\bf R}\!-\!\frac{\bf r}{2})}
  \Big )
\\
  \label{fieldDyn2_0i}
  i\hbar \frac{d\varphi_{0i}({\bf R, r})}{dt}
  &=&
  \Big (
    -\frac{\hbar^2 \nabla^2_R}{4m}
    -\frac{\hbar^2 \nabla^2_r}{m}
    +\underline{U_{aa}({\bf r})}
  \Big )
   \varphi_{0i}({\bf R, r})
\end{eqnarray}

where the one-body modes are coupled to the two-body modes through the coupling factors:
\begin{eqnarray}
  g_{00}({\bf x}) &=& \int\!\! d^3\! {\bf y}~\underline{U_{aa}({\bf x}\!\!-\!\!{\bf y})} \varphi_{00}({\bf x},{\bf y})
\\
  g_{0i}({\bf x}) &=& \int\!\! d^3\! {\bf y}~\underline{U_{aa}({\bf x}\!\!-\!\!{\bf y})} \varphi_{0i}({\bf x},{\bf y})
\\
  g_M({\bf R}) &=& \sqrt2 \int\!\!d^3\!{\bf r}~ \varphi_M^*({\bf r}) H_{int}({\bf R},{\bf r})\varphi_{00}({\bf R},{\bf r})
\\
  g'_M({\bf x}) &=& \sqrt 2  \int\!\! d^3\!{\bf y}~\underline{\varphi^*_M({\bf x}\!-\!{\bf y}) H_{int}({\bf x},{\bf y}) }
\end{eqnarray}

Note that the non-condensed modes $\Psi_i$ as well as the pairs $\varphi_{0i}$ involving a non-condensed atom are not directly affected by the molecular condensate. Thus, our equations cannot present any ``rogue photodissociation''\cite{kostrun00,javanainen02}, enabling molecules to dissociate towards non-condensed modes. The rogue photodissociation may be found either in the fluctuations of the molecular field or in the pairs $\varphi_{ij}$ involving two non-condensed atoms, which we have both neglected. However, rogue photodissociation has been treated thus far without the molecular fluctuations, and using the anomalous average $\langle \hat{a}_{\bf k} \hat{a}_{\bf -k}\rangle$ to describe non-condensed pairs. In our own approach, this average is rather related to the condensed pairs at short distances (see Eq. (\ref{Psi00})). We plan to clarify this point in a future work. In the rest of this paper, we will assume that we are in situations where the rogue photodissociation does not play an important role.\\

In order to decouple the one-body dynamics from the two-body dynamics, the pair wave functions must be quasi-stationary and follow adiabatically the one-body wave functions. In other words, the characteristic time of evolution for the pairs $T_{pairs}$ (see Eq. (\ref{T_pairs})) must be very short compared with the characteristic time of evolution $T_{cond}$ for the condensates:
\begin{equation}
\frac{T_{pairs}}{T_{cond}} \ll 1
\label{TimeComparison}
\end{equation}

In that case, we can solve Eqs. (\ref{fieldDyn2_00}),(\ref{fieldDyn2_0i}) formally using the Green's functions introduced in Eq. (\ref{varphi_0}):
\begin{eqnarray*}
\varphi_{00}({\bf R, r})&=& \varphi^{(0)}({\bf r}) 
\\&&-\int d^3{\bf R'}d^3{\bf r'}~ G({\bf R', r', R, r})
\sqrt2 H^*_{int}({\bf R',r'}) \varphi_M({\bf r'})
   \left (
    \frac{\Psi_M({\bf R'})}{\Psi_0({\bf R'}\!+\!\frac{\bf r'}{2})\Psi_0({\bf R'}\!-\!\frac{\bf r'}{2})}
  \right )
  \\
  ~& \approx & \varphi^{(0)}({\bf r}) - \int d^3{\bf r'}~ g({\bf r', r})
   \sqrt2 H^*_{int}({\bf r'}) \varphi_M({\bf r'})
   \frac{\Psi_M({\bf R})} {\Psi^2_0({\bf R})}
\\
\varphi_{0i}({\bf R, r})& =& \varphi^{(0)}({\bf r})
\end{eqnarray*} 

We can then eliminate $\varphi_{00}$ and $\varphi_{0i}$ in Eqs. (\ref{fieldDyn2_0}),(\ref{fieldDyn2_i}),(\ref{fieldDyn2_M}), and using the properties (\ref{varphi_0}),(\ref{scatteringLength}) of $\varphi^{(0)}$, we find:
\begin{eqnarray}
\label{MeanField20}
i\hbar \frac{d \Psi_0}{dt} &=&
\left (
  -\frac{\hbar^2 \nabla^2}{2m} + V + g\big(|\Psi_0|^2 + \sum_{i\ne0} 2|\Psi_i|^2\big)
\right ) \Psi_0 ~+~ w^* \Psi^*_0 \Psi_M
\\
\label{MeanField2i}
i\hbar \frac{d \Psi_i }{dt} &=&
\left (
  -\frac{\hbar^2 \nabla^2}{2m} + V + 2g|\Psi_0|^2
\right ) \Psi_i
\\
\label{MeanField2M}
i\hbar \frac{d \Psi_M }{dt} &=&
\left (
  -\frac{\hbar^2 \nabla^2}{4m} + 2 V + \hbar \delta + E_{self}
\right ) \Psi_M
~~+~~ \frac{1}{ 2} w \Psi_0^2
\end{eqnarray}

where $E_{self} = \int \frac{|\langle\varphi_M|H_{int}|\varphi^{(E)}\rangle|^2}{E} \rho(E)dE$ is the self-energy of the molecules: it is an energy shift caused by the interaction with the open channel \cite{fedichev96,bohn99,gerton01,mckenzie02}. We may include this energy shift in the detuning $\delta$. Note that Eq. (\ref{MeanField2i}) is conservative, so that the non-condensed modes may be omitted if they are initially negligible. Finally, we retrieve the coupled Gross-Pitaevski\u{\i} equations (\ref{MeanField10},\ref{MeanField1m}).

One can see in Eqs. (\ref{MeanField20}),(\ref{MeanField2M}) that the typical time $T_{cond}$ for the Rabi oscillations between $\Psi_0$ and $\Psi_M$ is given by $\hbar/(w\sqrt\rho)$, where $\rho$ is the typical density of the system while the interaction $w$ has been defined in Eq. (\ref{eq:w}). The condition (\ref{TimeComparison}) is thus equivalent to:
\begin{equation}
\label{condition}
\sqrt{2\rho} \langle \varphi_M | \varphi^{(0)} \rangle \ll 1
\end{equation}
 
Note that this condition does not depend on the intensity of the coupling $\bar H_{int}$, but keep in mind that it has been derived in a perturbative way. More generally, mean-field equations such as (\ref{MeanField20}),(\ref{MeanField2M}) should be valid as long as the time scale $T_{pairs}$, associated with the two-body dynamics, is short compared with the time scale $T_{cond}$ of the one-body dynamics. In other cases, one might need to use the more general Eqs. (\ref{fieldDyn_0}),(\ref{fieldDyn_M}) and treat the two-body dynamics explicitly.

\subsection{Comparison with other models}
These general equations can be compared with the models developed by Holland et al. \cite{kokkelmans02} and K\"{o}hler et al. \cite{kohler02}. In the Hartree-Fock-Bogoliubov model of Holland et al., anomalous and normal correlation functions are used to go beyond the coupled Gross-Pitaevski\u{\i} equations: when these correlation functions are negligible, their model also leads to the coupled Gross-Pitaevski\u{\i} equations. We therefore believe that the anomalous correlation is in fact related at large distances to the fluctuations around the stationary perturbed pair wave function $\Phi_{00}$. It is not true, however, at short distances, since their model is built with contact interactions. This is why a renormalization procedure is needed to recover the physics from short distances.
  
In the model of K\"{o}hler et al, the system is described by a single non-linear non-Markovian Schr\"{o}dinger equation which is not obviously related to the coupled Gross-Pitaevski\u{\i} equations. However, the cumulant approach at the basis of their model should be mathematically very close to our pair wave function approach. The main differences are that we factorize the correlations and consider the two channels explicitly: what they note $\Psi({\bf x})\Psi({\bf y})+\Phi({\bf x},{\bf y})$ is a condensed pair wave function with two implicit components corresponding to the pair wave functions $\Phi_{00}$ and $\Phi_M$. As our model has an explicit connection to the coupled Gross-Pitaevski\u{\i} equations, we believe it is also the case for the model of K\"{o}hler. Our model therefore clarifies the relation between the existing models and gives a simple image in terms of one-body and two-body wave functions.

\section{APPLICATION TO PHOTOASSOCIATION IN A BEC}

\begin{figure}[t]
  \begin{center}
    \includegraphics[width=8cm]{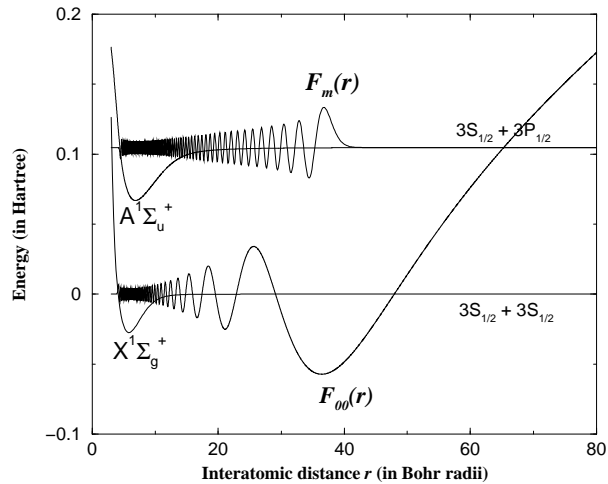}
  \end{center}
  \caption{
    Representation of the radial wave functions in the photoassociation experiment of Ref.\cite{mckenzie02}. The function in the open channel and the bound wave function were computed using respectively the Numerov-Cooley method \cite{cooley61} and the Fourier Grid Method \cite{slava99}. Note that in the text the total wave functions $\varphi_{00}({\bf r})=F_{00}(r)/{\sqrt{4\pi} r}$ and $\varphi_M({\bf r})=F_M(r)/{\sqrt{4\pi} r}$ are considered.}
  \label{fig:functions}
\end{figure}

The condition (\ref{condition}) shows that the one-body mean-field approach holds for sufficiently low densities, contrary to what has been suggested in ref. \cite{drummond02}. Actually, in the case of photoassociation, the condition (\ref{condition}) is usually achieved in the current experiments where the condensate is always dilute enough ($\rho \sim 10^{20} m^{-3}$), and the Franck-Condon factor $\langle \varphi_M | \varphi^{(0)} \rangle$ is sufficiently low ($\sim 10^{-14} m^{3/2}$). We give in Table I typical values. We first have considered the photoassociation experiment in a Na condensate reported in \cite{mckenzie02}, where the level $v=135$ of Na$_2$ $0_u^+(3S+3P_{1/2})$ (corresponding to $E_{bound}\sim 49 \  cm^{-1}$) is populated for various laser intensities. The computed radial wavefunctions are represented in Fig.(\ref{fig:functions}) : although the Franck-Condon factor is favourable for photoassociation, we find for a density of $4\times 10^{20} \ m^{-3}$ a ratio  $T_{pairs} / T_{cond} < 10^{-3}$. It is not easy to increase the transition moment by three orders of magnitude. Photoassociating towards molecular levels closer to the dissociation limit, for instance $v\ =\ 163$ (corresponding to $E_{bound}\sim 4~cm^{-1}$ and an outer turning point located at $r \sim 85~au$), would increase the Franck-Condon factor by less than one order of magnitude. Similar conclusions on the relative time scales can be drawn for a $Rb$ condensate, in the conditions described in Ref.\cite{drummond02}. We then expect that in most realistic cases the condition $T_{pairs} \ll T_{cond}$ will indeed be verified.

\begin{table}[t]
  \setlength{\tabcolsep}{0.2cm}
  \begin{tabular}{|c|c|c|c|c|c|c|}
  \hline
      ~ &
      Intensity &
      $E_{bound}$ &
      $T_{pairs}$ &
      $T_{cond}$ &
      $T_{spont}$ &
      $\sqrt{2\rho}\langle\varphi_m|\varphi^{(0)}\rangle$
  \\
  \hline\hline
      $^{23}Na$ \cite{mckenzie02} &
      $100~W/cm^{2}$ &
      $49~cm^{-1}$ &
      $8.4\times10^{-10} s$ &
      $1.1\times10^{-6} s$ &
      $8.4\times10^{-9} s$ &
      $7.4\times10^{-4}$
  \\
  \cline{2-5}
      ~ &
      $1~kW/cm^{2}$ &
      $49~cm^{-1}$ &
      $2.6\times10^{-10} s$ &
      $3.5\times10^{-7} s$ &
      ~ &
      ~
  \\
  \cline{2-5} \cline{7-7}
      ~ &
      $1~kW/cm^{2}$ &
      $4~cm^{-1}$ &
      $2.6\times10^{-10} s$ &
      $3.5\times10^{-7} s$ &
      ~ &
      $4.5\times10^{-3}$
  \\
  \hline
      $^{85}Rb$ \cite{drummond02} &
      ~ &
      ~ &
      $\sim 10^{-11} s$ &
      $4.8\times10^{-8} s$ &
      $1.4\times10^{-8} s$ &
      $\sim 3.10^{-4}$
  \\
  \hline
  \end{tabular}
  \caption{\label{values} Some typical time scales for the photoassociation of a condensate.}
\end{table}

There is another time scale, however, since the molecular state is not stable and decays by spontaneous emission. We can treat this spontaneous emission phenomenologically within the framework of our equations by adding a loss term $-i\hbar\frac{\gamma}{2}\Psi_M$ in the equation of the molecular field (\ref{fieldDyn2_M}). To simplify, we will consider a homogeneous system ($V=0$).

For sufficiently high intensities, $\bar H_{int} \gg \hbar\gamma$, ie $T_{pairs}\ll T_{spont}$, and the pairs have the time to oscillate between the two channels before spontaneous emission takes place. Eliminating the pair dynamics, we are left with a set of coupled Gross-Pitaevski\u{\i} equations:
\begin{eqnarray}
i\hbar \frac{d \Psi_0}{dt} &~=~&
  g|\Psi_0|^2 \Psi_0 ~~+~~ w^* \Psi^*_0 \Psi_M
\\
i\hbar \frac{d \Psi_M}{dt} &~=~&
\hbar (\delta - i\gamma/2) \Psi_M~~+~~ \frac{1}{ 2} w \Psi_0^2
\end{eqnarray}
Spontaneous emission is then usually faster than the dynamics of the condensates ($\hbar\gamma \gg w\sqrt{\rho}$), so that we can eliminate the molecular field adiabatically:
\begin{equation}
  \Psi_M = -\frac{w}{2\hbar(\delta-i\gamma/2)} \Psi^2_0
\end{equation}
We obtain a simple rate equation for the condensate density $|\Psi_0|^2$:
\begin{equation}
  \label{rateEquation}
  \frac{d|\Psi_0|^2}{dt} = -K |\Psi_0|^4
\end{equation}
with a rate $K=\frac{K_0}{1+\left(2\delta/\gamma\right)^2}$ and an on-resonance rate:
\begin{equation}
  \label{rate}
K_0=\frac{4}{\gamma}|\frac{1}{\hbar}\langle \varphi_M | H_{int} | \varphi^{(0)} \rangle|^2
\end{equation}

This is exactly the rate one can derive from a perturbative treatment of the two-atom equations (\ref{RelativeDyn1}),(\ref{RelativeDyn2}) \cite{bohn99}, assuming the presence of the condensate has only two effects: very low collision energies in the open channel and no need to symmetrize $\Phi_A$ since condensed atoms are in the same state. This last effect reduces the rate by a factor of 2 relative to a non-condensed gas \cite{Stoof89,kostrun00}, and was accounted for in our model through the proper symmetrization of the pair wave functions (compare Eqs. (\ref{reduced_ii}) and (\ref{reduced_ij})).
Thus, a cautious many-body treatment of the photoassociation in a BEC eventually yields the same prediction as two-atom theories. This indicates that the physics of the problem lies essentially in the pairs of atoms. This result has been confirmed by the photoassociation experiment in a condensate of sodium atoms reported by McKenzie {\it et al} \cite{mckenzie02}. The laser intensities used in this experiment (from $\sim 0.1$ to $1~kW.cm^{-2}$) are such that we are in the case considered here, where $T_{pairs} \ll T_{spont}$ (see Table~\ref{values}). They found that the condensate is indeed locally depleted according to Eq. (\ref{rateEquation}), with a rate $K_0$ proportional to the intensity, such that ${dK_0}/{dI}=3.5 \times 10^{-10} s^{-1}W^{-1}cm^{2}$. This is in agreement with the theoretical rate given by Eq. (\ref{rate}). Our own calculation, using numerical computations of stationary wave functions $\varphi_M$ and $\varphi^{(0)}$ (see Fig.\ref{fig:functions})), leads to ${dK_0}/{dI}=3.3 \times 10^{-10} s^{-1}W^{-1}cm^{2}$, in agreement with the calculation given in \cite{mckenzie02}.

 For smaller intensities, $\bar H_{int}\ll\hbar\gamma$, and the spontaneous emission is the fastest phenomenon: the formed molecules decay almost instantly, so that we can eliminate them adiabatically. Setting ${d\Psi_M}/{dt}=0$ in equation (\ref{fieldDyn2_M}) yields:
\begin{equation}
  \Psi_M = -\frac{\langle \varphi_M | H_{int} | \varphi_{00} \rangle}{\sqrt2\hbar(\delta-i\gamma/2)} \Psi^2_0
\end{equation}
So that we are left with:
\begin{eqnarray}
  i\hbar \frac{d \Psi_0}{dt} &=& \Big( \int\!d^3\!r~J[\varphi_{00}]({\bf r}) \Big) |\Psi_0|^2\Psi_0
\\
  i\hbar \frac{d\varphi_{00}}{dt}({\bf r})
  &=&
   -\frac{\hbar^2}{m}\nabla^2 \varphi_{00}({\bf r}) + J[\varphi_{00}]({\bf r})
\end{eqnarray}

where $J[\varphi_{00}]({\bf r}) = U_{aa}({\bf r})\varphi_{00}({\bf r}) - H^*_{int}({\bf r})\varphi_M({\bf r})\frac{\langle \varphi_M | H_{int} | \varphi_{00} \rangle}{\hbar(\delta - i\gamma/2)} $. The situation is the following: the formation of excited molecules creates a hole at short distances in the pair distribution $\varphi_{00}$, through the imaginary part of $J$. This hole is refilled with waves coming from larger distances, while the whole condensate is being depleted. Although this picture is quite different from the higher-intensity regime, we believe it leads essentially to the same rate of depletion, as we do not expect to see new behaviour emerging from lower intensities.\\

\section{CONCLUSION}

We have generalized the pair wave function approach, introduced by Cherny for an atomic condensate, to a non-homogeneous Bose system with two coupled channels, such as a mixed atomic-molecular condensate. The treatment of pair correlations is performed using the physical potentials both in the atomic and the molecular channel. We have shown that the pair wave functions regularise the mean-field terms in the equations without requirement of  any renormalization procedure. This has enabled us to derive the coupled Gross-Pitaevski\u{\i} equations on firm grounds and to assess their range of validity by comparing characteristic times associated to the pair dynamics and to the condensate dynamics. \\

We have shown that in the case of photoassociation with a cw laser, at intensities allowing a perturbative model, the Gross-Pitaevski\u{\i} description is usually verified and even leads to the rates predicted simply from the photoassociation probability of two colliding atoms. Nevertheless, we believe that the two-body mean-field equations might be necessary in other situations with a non-trivial time dependence, such as photoassociation with chirped laser pulses. Previous works \cite{kokkelmans02,mackie02,kohler02} have already shown the importance of long-range correlations for Feshbach resonance induced by a time-dependent magnetic field. Future work will clarify the possible influence of short range correlations in these time-dependent situations.\\

\section{ACKNOWLEDGMENTS}

The authors would like to thank E. Tiesinga, M. Mackie, S. Kokkelmans, R. Kosloff and P. Pellegrini for helpful discussions.

\appendix
\section{MODES AND PAIRWAVE FUNCTIONS}

In this appendix we generalize the ideas of Cherny \cite{cherny00} to a non-homogeneous Bose gas. To simplify the notations, the field operator $\hat{\psi}_a$ will be noted $\hat{\psi}$.

\subsection{Definition of the pair wave functions}

In second quantization, the one-body and two-body density matrices of the system are defined by:
\begin{eqnarray}
  \label{DefOneBodyDensity}
  \rho_1({\bf x},{\bf y}) &\equiv& \langle \hat{\psi}^\dag({\bf x}) \hat{\psi}({\bf y}) \rangle
  \\
  \rho_2({\bf x}, {\bf y}, {\bf y'}, {\bf x'}) &\equiv&
  \frac{1}{2!} \langle \hat{\psi}^\dag({\bf x}) \hat{\psi}^\dag({\bf y}) \hat{\psi}({\bf y'}) \hat{\psi} ({\bf x'}) \rangle
\end{eqnarray}

As hermitian matrices, they can be diagonalised in an orthonormal basis:
\begin{eqnarray}
  \label{OneBodyDensity}
  \rho_1({\bf x},{\bf y}) &=& 
  \sum_{i} n_i \psi_i^*({\bf x}) \psi_i({\bf y})
  \\
  \rho_2({\bf x}, {\bf y}, {\bf y'}, {\bf x'}) &=&
  \sum_{\nu} m_{\nu} \phi_{\nu}^*({\bf x,y}) \phi_{\nu}({\bf x',y'})
\end{eqnarray}

This means that there are $n_i$ atoms in the one-body mode (or wave function) $\psi_i$, and there are $m_{\nu}$ pairs of atoms in the two-body mode (or pair wave function) $\phi_{\nu}$. All these wave functions are normalised to unity, but we can alternatively define the functions $\Psi_i = \sqrt n_i \psi_i$ and $\Phi_{\nu} = \sqrt{m_{\nu}} \phi_{\nu}$, which are normalised to the number of atoms or pairs of atoms. Note that in a state where the total number of atoms is $n$, we have $\sum_i n_i = n$ and $\sum_\nu m_\nu = \frac{n(n-1)}{2}$.
This may be seen as a consequence of the relations:
\begin{eqnarray}
  \int\!d^3\!x~ \langle \hat{\psi}^\dag({\bf x}) \hat{\psi}({\bf x}) \rangle &=& n
  \\
  \label{relation2-1}
  \int\!d^3\!y~ \langle \hat{\psi}^\dag({\bf x}) \hat{\psi}^\dag({\bf y}) \hat{\psi}({\bf y}) \hat{\psi} ({\bf x'}) \rangle
  &=& (n-1)\langle \hat{\psi}^\dag({\bf x}) \hat{\psi}({\bf x'}) \rangle
\end{eqnarray}

We can use the one-body modes $\psi_i$ to expand the field operator: $\hat{\psi}({\bf x}) = \sum_i \hat{a}_i \psi_i({\bf x})$. Using (\ref{DefOneBodyDensity}) and (\ref{OneBodyDensity}), we find the relation $\langle \hat{a}^\dag_i \hat{a}_j \rangle = n_i \delta_{ij}$.
 
When there is no interaction between the atoms, the atoms are decorrelated and the pair wave functions $\phi_{\nu}$ are simply symmetrized products of one-body wave functions:
\begin{eqnarray}
  \phi_{ii}({\bf x},{\bf y}) &=& \psi_i({\bf x})\psi_i({\bf y})
  \\
  \phi_{ij}({\bf x},{\bf y}) &=& \frac{\psi_i({\bf x})\psi_j({\bf y}) + \psi_j({\bf x})\psi_i({\bf y})}{\sqrt 2} ~~( i < j )
\end{eqnarray}

$\phi_{ii}$ is the wave function for two atoms in the same mode $\psi_i$, and $\phi_{ij}$ is the wave function for two atoms in the modes $\psi_i$ and $\psi_j$. From counting arguments, we have $m_{ii} = \frac{n_i(n_i-1)}{2}$ pairs in the mode $\phi_{ii}$ and $m_{ij} = n_i n_j$ pairs in the mode $\phi_{ij}$. One can check that this pair distribution does indeed satisfy the relation (\ref{relation2-1}).

\begin{figure}[t]
  \begin{center}
    \includegraphics[width=6cm]{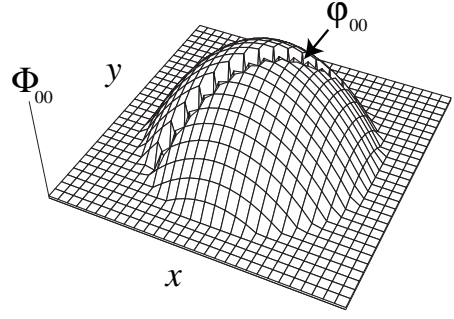}
  \end{center}
  \caption{
    Schematic representation of the macroscopic wave function for condensate pairs in a one dimensional system: $\Phi_{00}(x,y) = \frac{1}{\sqrt2}
\Psi_0(x)\Psi_0(y)\varphi_{00}(x,y)$. The reduced pair wave function $\varphi_{00}$ gives the correlation at short distance $|x-y|$ created by the scattering of the two atoms.
  }
  \label{fig:pair}
\end{figure}

In the presence of interactions, one might expect extra terms needed to describe the correlations at short interatomic distances $|{\bf x}-{\bf y}|$ (see Fig.\ref{fig:pair}):
\begin{eqnarray}
  \label{general_psi_ii}
  \phi_{ii}({\bf x},{\bf y}) &\propto& \psi_i({\bf x})\psi_i({\bf y}) + \phi'_{ii}({\bf x},{\bf y})
  \\
  \label{general_psi_ij}
  \phi_{ij}({\bf x},{\bf y}) &\propto&  \frac{\psi_i({\bf x})\psi_j({\bf y}) + \psi_j({\bf x})\psi_i({\bf y})}{\sqrt 2} + \phi'_{ij}({\bf x},{\bf y})
\end{eqnarray}

These correlations $\phi'_{ij}$ are expected to vanish for interatomic distances larger than, say, the range $r_0$ of the interaction. The pair wave functions are thus asymptotically decorrelated; this follows from the principle of correlation weakening\cite{cherny00}. If the interaction supports bound states, one might also expect bound-pair wave functions $\phi_\nu$, vanishing entirely for interatomic distances larger than $r_0$.

As the functions $\phi'_{ij}$ will scale like the functions $\psi_i$ and $\psi_j$, it can be useful to express the correlations with dimensionless functions, which we refer to as ``reduced pair wave functions'':
\begin{eqnarray}
  \label{reduced_ii}
  \phi_{ii}({\bf x},{\bf y}) &\propto& \psi_i({\bf x})\psi_i({\bf y})\varphi_{ii}({\bf x},{\bf y})
  \\
  \label{reduced_ij}
  \phi_{ij}({\bf x},{\bf y}) &\propto&  \frac{\psi_i({\bf x})\psi_j({\bf y}) + \psi_j({\bf x})\psi_i({\bf y})}{\sqrt 2}
  \varphi_{ij}({\bf x},{\bf y})
\end{eqnarray}
 
with $\varphi_{ij}({\bf x},{\bf y}) \to 1$ when $|{\bf x}\!-\!{\bf y}| \gg r_0$.

\subsection{Pair wave functions in a condensate with U(1) symmetry breaking}

When condensation occurs, most of the atoms are in the same quantum state, say $\phi_0$. This means $n_0 \gg \sum_i n_i$. Accordingly, most of the pairs are in the same quantum state $\psi_{00}$, corresponding to pairs of condensed atoms. A way of treating this situation, related to the Bogoliubov ansatz,  is to assume a breaking of the $U(1)$ symmetry\cite{leggett01}: the system is then in a coherent state $|\Omega\rangle$ satisfying $\hat{a}_0|\Omega\rangle = \alpha_0 |\Omega\rangle$. From the relation $\langle \hat{a}^\dag_i \hat{a}_j\rangle = n_i \delta_{ij}$, we obtain $\alpha_0 = \sqrt{n_0}$ and $\langle \hat{a}_i\rangle = 0 ~(i \ne 0)$, so that the field operator has a non-zero expectation value $\langle \hat{\psi} \rangle=\Psi_0$. The deviations from the mean $\hat{\theta} = \Psi_0 - \hat{\psi}$ are the quantum fluctuations.
This implies that the number of atoms now fluctuates and the relation (\ref{relation2-1}) derived for a fixed-number state is no more valid. However, this prescription is very useful to derive expressions for the one-body and two-body modes in terms of the quantum fluctuations.
Indeed, if we expand the two-body density matrix in terms of $\hat{\theta}$:
\begin{widetext}
\begin{eqnarray}
  \rho_2({\bf x},{\bf y},{\bf y'},{\bf x'})
  &=& \frac{1}{2} \Big \{
  \Psi^*_0({\bf x})\Psi^*_0({\bf y})\Psi_0({\bf y'})\Psi_0({\bf x'})
  + \Psi^*_0({\bf x})\Psi^*_0({\bf y})\langle \hat{\theta}({\bf y'}) \hat{\theta}({\bf x'}) \rangle 
  \\
  &+& \langle \hat{\theta}({\bf x}) \hat{\theta}({\bf y}) \rangle^* \Psi_0({\bf y'})\Psi_0({\bf x'}) \nonumber
  \\
  &+&
  \Psi_0^*({\bf x})\Psi_0({\bf x'})\langle \hat{\theta}^\dag\!({\bf y}) \hat{\theta}({\bf y'}) \rangle
  + \Psi_0^*({\bf x})\Psi_0({\bf y'})\langle \hat{\theta}^\dag\!({\bf y}) \hat{\theta}({\bf x'}) \rangle
  \\
  &+& \Psi_0^*({\bf y})\Psi_0({\bf x'})\langle \hat{\theta}^\dag\!({\bf x}) \hat{\theta}({\bf y'}) \rangle
  + \Psi_0^*({\bf y})\Psi_0({\bf y'})\langle \hat{\theta}^\dag\!({\bf x}) \hat{\theta}({\bf x'}) \rangle
  \nonumber
  \\
  &+&
  \Psi^*_0({\bf x}) \langle \hat{\theta}^\dag({\bf y}) \hat{\theta}({\bf y'}) \hat{\theta}({\bf x'}) \rangle
  + \Psi^*_0({\bf y}) \langle \hat{\theta}^\dag({\bf x}) \hat{\theta}({\bf y'}) \hat{\theta}({\bf x'}) \rangle
  \\
  &+& \Psi_0({\bf y'}) \langle \hat{\theta}^\dag({\bf x}) \hat{\theta}^\dag({\bf y}) \hat{\theta}({\bf x'}) \rangle
  + \Psi_0({\bf x'}) \langle \hat{\theta}^\dag({\bf x}) \hat{\theta}^\dag({\bf y}) \hat{\theta}({\bf y'}) \rangle
  \nonumber
  \\
  &+&  
  \langle \hat{\theta}^\dag({\bf x}) \hat{\theta}^\dag({\bf y}) \hat{\theta}({\bf y'}) \hat{\theta}({\bf x'})\rangle
  \Big \}
 \end{eqnarray}
 \end{widetext}
 
We see that it can be factorized as follows:
 \begin{eqnarray}
 \label{fourFields}
  \rho_2({\bf x},{\bf y},{\bf y'},{\bf x'})
  = 
  \sum_{i\le j} \Phi^*_{ij}({\bf x},{\bf y})\Phi_{ij}({\bf x'},{\bf y'})
  + \sum_{\nu} \Phi^*_{\nu}({\bf x},{\bf y})\Phi_{\nu}({\bf x'},{\bf y'})
\end{eqnarray}

where:
\begin{eqnarray}
\label{Psi00}
\Phi_{00}({\bf x},{\bf y}) &=& 
\frac{\langle \hat{\psi}({\bf x})\hat{\psi}({\bf y}) \rangle}{\sqrt2} =
\frac{\Psi_0({\bf x})\Psi_0({\bf y})}{\sqrt2} + \sum_{i\ne0,j\ne0} \frac{\langle \hat{a}_i\hat{a}_j\rangle}{\sqrt2}\psi_i({\bf x})\psi_j({\bf y})
\\
\Phi_{0i}({\bf x},{\bf y}) &=& \frac{\Psi_0({\bf x})\Psi_i({\bf y}) + \Psi_i({\bf x})\Psi_0({\bf y})}{\sqrt2}
+ \sum_{j\ne0, k\ne0} \frac{\langle \hat{a}^\dag_i \hat{a}_j \hat{a}_k \rangle}{\sqrt{2\langle \hat{a}^\dag_i\hat{a}_i\rangle}}
\psi_j({\bf x})\psi_k({\bf y})
\end{eqnarray}

are respectively the pair wave function for two condensed atoms, and the pair wave function for a condensed atom and a non-condensed atom in the mode $\Psi_i$. There is no simple expression for the pair wave function $\Phi_{ij}$ of two non-condensed atoms, nor the bound-pair wave functions $\Phi_{\nu}$, since these are contained in the term $\langle \hat{\theta}^\dag({\bf x}) \hat{\theta}^\dag({\bf y}) \hat{\theta}({\bf y'}) \hat{\theta}({\bf x'})\rangle$. If we assume that there is no interaction between the non-condensed atoms (which is the case in the usual Hartree-Fock-Bogoliubov approach), we may use Wick's theorem to expand this term:
\begin{eqnarray}
  \langle \hat{\theta}^\dag({\bf x}) \hat{\theta}^\dag({\bf y}) \hat{\theta}({\bf y'}) \hat{\theta}({\bf x'})\rangle ~ & = &
  ~~\langle \hat{\theta}^\dag({\bf x}) \hat{\theta}^\dag({\bf y}) \rangle \langle \hat{\theta}({\bf y'}) \hat{\theta}({\bf x'})\rangle \nonumber
\\
~ & ~ &  
  +~\langle \hat{\theta}^\dag({\bf x}) \hat{\theta}({\bf x'}) \rangle \langle \hat{\theta}^\dag({\bf y}) \hat{\theta}({\bf y'}) \rangle
\nonumber  
\\  
~ & ~ &
+~ \langle \hat{\theta}^\dag({\bf x}) \hat{\theta}({\bf y'}) \rangle \langle \hat{\theta}^\dag({\bf y}) \hat{\theta}({\bf x'})\rangle
\end{eqnarray}

This yields:
\begin{eqnarray}
\Phi_{00}({\bf x},{\bf y}) &=& \frac{\Psi_0({\bf x})\Psi_0({\bf y})}{\sqrt2} + \sum_{i\ne0,j\ne0} \frac{\langle \hat{a}_i\hat{a}_j\rangle}{\sqrt2}\psi_i({\bf x})\psi_j({\bf y})
\\
\Phi_{0i}({\bf x},{\bf y}) &=& \frac{\Psi_0({\bf x})\Psi_i({\bf y}) + \Psi_i({\bf x})\Psi_0({\bf y})}{\sqrt2}
\\
\Phi_{ii}({\bf x},{\bf y}) &=& \frac{\Psi_i({\bf x})\Psi_i({\bf y})}{\sqrt2}
\\
\Phi_{ij}({\bf x},{\bf y}) &=& \frac{\Psi_i({\bf x})\Psi_j({\bf y}) + \Psi_j({\bf x})\Psi_i({\bf y})}{\sqrt2}
\\
\Phi_{\nu}({\bf x},{\bf y}) &=& 0
\end{eqnarray}

As expected in this case, only the condensed pairs are correlated, through the anomalous correlation $\langle \hat{\theta}({\bf x}) \hat{\theta}({\bf y})\rangle = \sum_{i\ne0,j\ne0} \langle \hat{a}_i\hat{a}_j\rangle \psi_i({\bf x})\psi_j({\bf y})$. This means the condensed atoms interact only with each other, while the non-condensed atoms are treated as an ideal gas. However, this is an approximation, since all the atoms feel the interaction. As we shall see, we actually need the correlations in order to regularise mean-field terms appearing in the equations of motion for the one and two-body modes.

\subsection{Evolution of one-body modes}

The evolution of the condensate mode is simply obtained by taking the average of the equation of motion for the field operator (\ref{operatorDyn_a}). Doing this, we encounter the average $\langle \hat{\psi}^\dag({\bf x}) \hat{\psi}({\bf y}) \hat{\psi}({\bf z}) \rangle$, which can be expanded and then factorized in terms of pair wave functions:
\begin{eqnarray}
  \langle \hat{\psi}^\dag({\bf x}) \hat{\psi}({\bf y}) \hat{\psi}({\bf z}) \rangle &=&
  ~~~\Psi_0^*({\bf x}) \langle \hat{\theta}({\bf y}) \hat{\theta}({\bf z}) \rangle
  + \Psi_0({\bf y}) \langle \hat{\theta}^\dag({\bf x}) \hat{\theta}({\bf z}) \rangle \nonumber
  \\ 
  &&+~ \Psi_0({\bf z}) \langle \hat{\theta}^\dag({\bf x}) \hat{\theta}({\bf y}) \rangle
  + \langle \hat{\theta}^\dag({\bf x}) \hat{\theta}({\bf y}) \hat{\theta}({\bf z}) \rangle
  \\
  &=&
  \label{threeFields}
  \sqrt2 \sum_i \Psi^*_i({\bf x}) \Phi_{0i}({\bf y},{\bf z})
\end{eqnarray} 

We also have to express the average $\langle \hat{\psi}^\dag_a({\bf y}) \underline{H_{int}^*({\bf x},{\bf y}) \hat{\psi}_b({\bf x}) \hat{\psi}_c({\bf y})}\rangle$ which can be simplified to $\Psi^*_0({\bf y}) \underline{H_{int}^*({\bf x},{\bf y})} \Phi_M({\bf x},{\bf y})$ if we assume that the atoms of the upper channel are paired in a purely coherent field $\Phi_M$. The equation of motion for the condensed mode is then:
\begin{eqnarray}
\label{Phi_0}
&&i\hbar \frac{d {\Psi}_0 ({\bf x})}{dt} =
\left (
  -\frac{\hbar^2 \nabla^2_x}{2m} + V({\bf x})
\right ) \Psi_0 ({\bf x}) \nonumber
\\
&&+~ \sqrt2 \int\!\! d^3\! {\bf y} \left ( \sum_i \Psi^*_i({\bf y}) \underline{U_{aa}({\bf x}\!\!-\!\!{\bf y})} \Phi_{0i}({\bf x},{\bf y})
+  \Psi^*_0({\bf y}) \underline{H_{int}^*({\bf x},{\bf y})} \Phi_M({\bf x},{\bf y}) \right )
\end{eqnarray}

We can then derive the evolution of the one-body density matrix for the non-condensed atoms $\rho'_1({\bf x},{\bf y}) \equiv \langle \hat{\psi}^\dag({\bf x})\hat{\psi}({\bf y}) \rangle - \Psi^*_0({\bf x})\Psi_0({\bf y}) = \sum_{i\ne0} \Psi^*_i({\bf x})\Psi_i({\bf y})$. Using Eqs. (\ref{operatorDyn_a}), (\ref{fourFields}), (\ref{fieldDyn_0}), and neglecting the pair wave functions $\Phi_{ij}$ for two non-condensed atoms, we get:
\begin{eqnarray}
&&i\hbar \frac{d {\rho'_1} ({\bf x},{\bf y})}{dt} ~=~ \langle \hat{\psi}^\dag({\bf x})i\hbar\frac{d\hat{\psi}({\bf y})}{dt} \rangle - \Psi^*_0({\bf x})i\hbar\frac{d\Psi_0({\bf y})}{dt}
- \{ {\bf x} \leftrightarrow {\bf y}\}^*
\nonumber
\\
&=&
\left (-\frac{\hbar^2\nabla^2_y}{2m} + V({\bf y}) \right) \rho'_1({\bf x},{\bf y})
\nonumber
\\
&+&
2\int\!\!d^3\!z~ \underline{U_{aa}({\bf y}\!\!-\!\!{\bf z})}
\left( 
    {\Phi'}^*_{00}({\bf x},{\bf z})\Phi_{00}({\bf z},{\bf y})
    + \sum_{i\ne0} \left( \frac{\Psi_0^*({\bf z}){\Psi_i}^*({\bf x})}{\sqrt2} + \Phi'_{0i}({\bf x},{\bf z}) \right)\Phi_{0i}({\bf z},{\bf y})
\right)
\nonumber
\\
&+&
2\int\!\!d^3\!z~ {\Phi'}^*_{00}({\bf x},{\bf z}) \underline{H^*_{int}({\bf y},{\bf z})} \Phi_M({\bf z},{\bf y})
- \{ {\bf x} \leftrightarrow {\bf y}\}^*
\end{eqnarray}

In the expression above, the correlation terms $\Phi'({\bf x},{\bf z})$ vanish for $|{\bf y}-{\bf x}|\gg r_0$. So when ${\bf x}$ and ${\bf y}$ are separated by macroscopic distances, we have:

\begin{eqnarray}
i\hbar \frac{d {\rho'_1} ({\bf x},{\bf y})}{dt} &\approx&
\left (-\frac{\hbar^2\nabla^2_y}{2m} + V({\bf y}) \right) \rho'_1({\bf x},{\bf y})
\nonumber
\\
&&+~2\int\!\!d^3\!z~ \underline{U_{aa}({\bf y}\!\!-\!\!{\bf z})}
  \sum_{i\ne0} \frac{\Psi_0^*({\bf z})\Psi_i^*({\bf x})}{\sqrt2}\Phi_{0i}({\bf z},{\bf y})
  - \{ {\bf x} \leftrightarrow {\bf y}\}^*
\end{eqnarray}

which can be written as:

\begin{multline}
\sum_{i\ne0}\Psi^*_i({\bf x}) i\hbar \frac{d \Psi_i({\bf y})}{dt} - \{ {\bf x} - {\bf y}\}^* \approx
\\
\sum_{i\ne0}\Psi^*_i({\bf x})
\left [
\left (-\frac{\hbar^2\nabla^2_y}{2m} + V({\bf y}) \right) \Psi_i({\bf y})
+
\sqrt2\int\!\!d^3\!z~ \underline{U_{aa}({\bf y}\!\!-\!\!{\bf z})}\Psi_0^*({\bf z})\Phi_{0i}({\bf z},{\bf y})
\right ] - \{ {\bf x} \leftrightarrow {\bf y}\}^*
\end{multline}

It is clear from this expression that the evolution of the non-condensed mode $\Psi_i$ is given by:
\begin{eqnarray}
\label{Phi_i}
i\hbar \frac{d \Psi_i({\bf x})}{dt} &\approx&
\left (-\frac{\hbar^2\nabla^2_x}{2m} + V({\bf x}) \right) \Psi_i({\bf x})
+
\sqrt2\int\!\!d^3\!y~ \underline{U_{aa}({\bf x}\!\!-\!\!{\bf y})}\Psi_0^*({\bf y})\Phi_{0i}({\bf y},{\bf x})
\end{eqnarray}

as long as its wavelength is greater than $r_0$. The interpretation of Eqs. (\ref{Phi_0}) and (\ref{Phi_i}) is straightforward. Every condensed atom evolving in the trap can collide with another condensed atom or with any non-condensed atoms, and it can also associate with another condensed atom to form a molecule. Similarly, every non-condensed atom evolving in the trap can collide with any condensed atom (we neglected the collisions with other non-condensed atoms). Note that, within the approximation of a purely coherent molecular field, the non-condensed modes are not directly affected by the formation of molecules.

\subsection{Evolution of the two-body modes}

The equation of motion for the pair wave function of condensed atoms follows from Eqs. (\ref{Psi00}) and (\ref{operatorDyn_a}):
\begin{eqnarray}
  &&i\hbar\frac{d\Phi_{00}}{dt}({\bf x},{\bf y})
  = \frac{1}{\sqrt2} \langle \hat{\psi}({\bf x}) i\hbar\frac{\hat{\psi}}{dt}(\bf y) \rangle + \{ {\bf x} \leftrightarrow {\bf y}\}
  \\
  &=& 
  \left (
    -\frac{\hbar^2 (\nabla^2_x\!+\!\nabla^2_y)}{2m} + V({\bf x})+ V({\bf y})
    +\underline{U_{aa}({\bf x\!-\!y})}
  \right )
  \Phi_{00}({\bf x},{\bf y})
  ~+~
  H^*_{int}({\bf x},{\bf y})\Phi_M({\bf x},{\bf y})
  \nonumber
\\
  \label{Phi_00_evolution}
  &&+~ \int\!\! d^3\! {\bf z}~ 
  \left ( 
    \underline{U_{aa}({\bf z}\!\!-\!\!{\bf y})} \sigma_2({\bf z},{\bf z},{\bf x},{\bf y})
    + \rho_1({\bf z},{\bf x}) H_{int}^*({\bf y},{\bf z}) \Phi_M({\bf y},{\bf z})
    + \{ {\bf x} \leftrightarrow {\bf y}\}
  \right )
\end{eqnarray}

with $\sigma_2({\bf z},{\bf z},{\bf x},{\bf y}) = \frac{1}{\sqrt2}\langle \hat{\psi}^\dag({\bf z})\hat{\psi}({\bf z})\hat{\psi}({\bf x})\hat{\psi}({\bf y}) \rangle$. Using Eq. (\ref{Phi_0}), one can then deduce the equation of motion for the reduced pair wave function $\varphi_{00}$ :
\begin{eqnarray}
  \label{varphi_00}
  &&i\hbar \frac{d \varphi_{00} ({\bf x}, {\bf y})}{dt} =
  \Big [ ~
    -\frac{\hbar^2 }{2m}(\nabla^2_x\!+\!\nabla^2_y)
    +\underline{U_{aa}({\bf x\!-\!y})} 
  \nonumber
  \\
  &&-\frac{\hbar^2 }{m} (\nabla\ln\!\Psi_0({\bf x})\cdot\nabla_x~+~\nabla\ln\!\Psi_0({\bf y})\cdot\nabla_y)
  \nonumber
  \\
  &&+~ \int\!\! d^3\! {\bf z}~
  \left (
  \underline{U_{aa}({\bf z}\!\!-\!\!{\bf y})} F({\bf x},{\bf y},{\bf z})
   ~+~
   G({\bf x},{\bf y},{\bf z}) \sqrt2 H_{int}({\bf x},{\bf z})\Phi_M({\bf x},{\bf z})
  + \{ {\bf x} \leftrightarrow {\bf y}\}
  \right )
  ~ \Big ]
  \varphi_{00}({\bf x},{\bf y})
  \nonumber
  \\
  &+& \sqrt2 H^*_{int}({\bf x},{\bf y})\frac{\Phi_M({\bf x},{\bf y})}{\Psi_0({\bf x})\Psi_0({\bf y})}
\end{eqnarray}

with $F({\bf x},{\bf y},{\bf z}) = 
    \frac{
      \langle \hat{\psi}^\dag({\bf z}) \hat{\psi}({\bf z}) \hat{\psi}({\bf x}) \hat{\psi}({\bf y}) \rangle
    }{
      \langle \hat{\psi}({\bf x}) \hat{\psi}({\bf y}) \rangle
    }
    -
    \frac{
      \langle \hat{\psi}^\dag({\bf z}) \hat{\psi}({\bf z}) \hat{\psi}({\bf y}) \rangle
    }{
      \langle \hat{\psi}({\bf y}) \rangle
    }$
and $G({\bf x},{\bf y},{\bf z}) = 
    \frac{
      \langle \hat{\psi}^\dag({\bf z}) \hat{\psi}({\bf y}) \rangle
    }{
      \langle \hat{\psi}({\bf x}) \hat{\psi}({\bf y}) \rangle
    }
    -
    \frac{
      \langle \hat{\psi}^\dag({\bf z}) \rangle
    }{
      \langle \hat{\psi}({\bf x}) \rangle
    }$.
These many-body terms are caused by the interaction with other atoms external to the pair considered. At sufficiently low densities, we can make two basic assumptions about the reduced pair wave function:
\begin{description}
\item[H1:] its extent $\sim r_0$ is much smaller than the typical length scale of the condensate
\item[H2:] it is not much influenced by remaining many-body terms like $F$ and $G$. The many-body effects on the pair dynamics are thus only caused by the mean fields $\Psi_i$, through the relations (\ref{reduced_ii}) and (\ref{reduced_ij}).  
\end{description}
Within these assumptions, Eq. (\ref{varphi_00}) reduces to:
\begin{equation}
  \label{varphi_00simplified}
  i\hbar \frac{d \varphi_{00} ({\bf x}, {\bf y})}{dt} =
  \Big (
    -\frac{\hbar^2 }{2m}(\nabla^2_x\!+\!\nabla^2_y)
    +\underline{U_{aa}({\bf x\!-\!y})} 
  ~ \Big )
  \varphi_{00}({\bf x},{\bf y})
  ~+~
  \sqrt2 H^*_{int}({\bf x},{\bf y})\frac{\Phi_M({\bf x},{\bf y})}{\Psi_0({\bf x})\Psi_0({\bf y})}
\end{equation}

The pair wave function $\Phi_{0i}$ for a condensed atom and a non-condensed atom in the mode $i$ is given by $\frac{1}{\sqrt2}\int\!d^3\!z~\Psi_i({\bf z})\langle \hat{\psi}^\dag({\bf z}) \hat{\psi}({\bf x}) \hat{\psi}({\bf y}) \rangle$, following from Eq. (\ref{threeFields}). Differentiating this expression and using the assumptions ${\bf H_1}$ and ${\bf H_2}$, we find for the reduced pair wave function $\varphi_{0i}$:
\begin{equation}
  \label{varphi_0isimplified}
  i\hbar \frac{d \varphi_{0i} ({\bf x}, {\bf y})}{dt} =
  \Big (
    -\frac{\hbar^2 }{2m}(\nabla^2_x\!+\!\nabla^2_y)
    +\underline{U_{aa}({\bf x\!-\!y})} 
  ~ \Big )
  \varphi_{0i}({\bf x},{\bf y})
\end{equation}
Within our approximations, a pair involving a non-condensed atom is not directly affected by the coupling between the channels. This may indicate that one actually needs the quantum fluctuations of the molecular field for these modes.

Under the assumption ${\bf H_1}$, we can simplify the equations (\ref{Phi_0}) and (\ref{Phi_i}) for the one-body modes:
\begin{eqnarray}
\label{Phi_0simplified}
i\hbar \frac{d {\Psi}_0}{dt} &=&
\Big (
  -\frac{\hbar^2 \nabla^2}{2m} + V
  + g_{00}|\Psi_0|^2 
  + \sum_{i\ne0} 2 g_{0i}|\Psi_i|^2 
\Big ) \Psi_0 + G_M \Psi^*_0  
\\
\label{Phi_isimplified}
i\hbar \frac{d \Psi_i}{dt} &=&
\Big (-\frac{\hbar^2\nabla^2}{2m} + V + 2 g_{0i}|\Psi_0|^2  \Big ) \Psi_i
\end{eqnarray}

with:
\begin{eqnarray}
g_{00}({\bf x}) &=& \int\!\!d^3\!y~ \underline{U_{aa}({\bf x\!-\!y})}\varphi_{00}({\bf x,y})
\\
g_{0i}({\bf x}) &=& \int\!\!d^3\!y~ \underline{U_{aa}({\bf x\!-\!y})}\varphi_{0i}({\bf x,y})
\\
G_M({\bf x}) &=& \sqrt2 \int\!\!d^3\!y~ \underline{H_{int}^*({\bf x},{\bf y})} \Phi_M({\bf x},{\bf y})
\end{eqnarray}

\begin{figure}[t]
  \begin{center}
    \includegraphics[width=3cm]{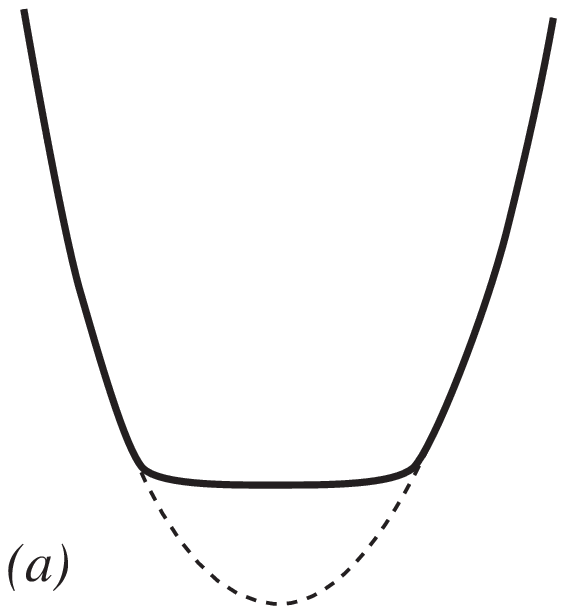}
    \includegraphics[width=3cm]{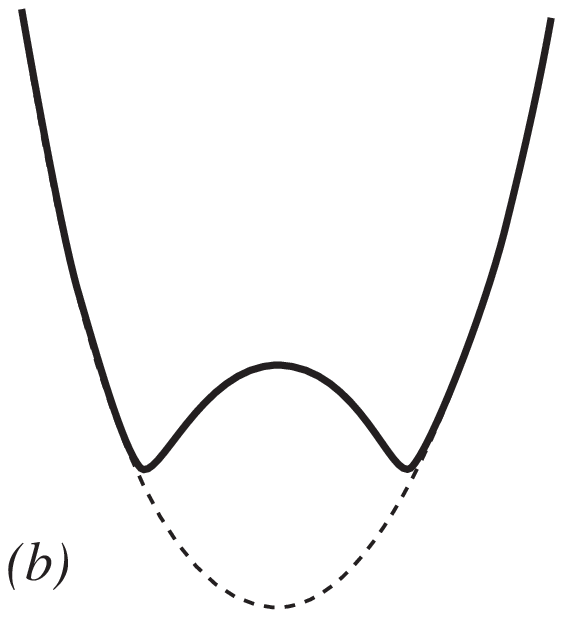}
  \end{center}
  \caption{
    Schematic representation of the effective potentials felt by $(a)$ the condensed atoms, $(b)$ the non-condensed atoms (for the first non-condensed modes), in the case of a positive scattering length.
  }
  \label{fig:traps}
\end{figure}

These factors are proportional to scattering amplitudes. Note that the amplitudes are twice larger for pairs of condensed and non-condensed atoms; this comes from the fact that the atoms are in different states, so that the symmetrization (\ref{general_psi_ij}) induces a factor of two relative to a pair of condensed atoms. As a consequence, the effective potentials felt by the condensed atoms and the non-condensed atoms are different. Fig.\ref{fig:traps} gives a schematic representation of these potentials at zero temperature equilibrium, where $g_{00}=g_{0i}=g=\frac{4\pi\hbar^2a}{m}$, with a positive scattering length $a$.

\end{document}